\documentclass[pra,tightenlines,secnumarabic,eqsecnum,12pt]{revtex4}

\usepackage{bm}

\usepackage{graphicx}

\def\vec#1{{\bm #1}}

\begin{document}

\title{I-concurrence and tangle for isotropic states}

\author{Pranaw Rungta}
\author{Carlton M.~Caves}
\affiliation{Department of Physics and Astronomy,
University of New Mexico, Albuquerque, NM 87131--1156, USA}

\date{2002 June~20}


\begin{abstract}
We discuss properties of entanglement measures called {\em
I-concurrence\/} and {\em tangle}.  For a bipartite pure state,
I-concurrence and tangle are simply related to the purity of the
marginal density operators.  The I-concurrence (tangle) of a
bipartite mixed state is the minimum average I-concurrence (tangle)
of ensemble decompositions of pure states of the joint density
operator.  Terhal and Vollbrecht [Phys.\ Rev.\ Lett.\ {\bf 85}, 2625
(2000)] have given an explicit formula for the entanglement of
formation of isotropic states in arbitrary dimensions.  We use their
formalism to derive comparable expressions for the I-concurrence and
tangle of isotropic states.
\end{abstract}

\maketitle

\section{Introduction}
\label{sec:intro} Entanglement among quantum systems is a distinctive
feature of quantum
mechanics~\cite{Einstein1935,Bell1964,Wootters2001} and an
indispensable ingredient in various kinds of quantum information
processing protocols \cite{Lo1998,Mnielsen2000}, e.g., quantum
teleportation.  It is desirable to have a general theory of
entanglement, but though important results have been obtained, a
general theory has proven elusive because of the complex nature of
entanglement for all but the very simplest joint quantum systems.
Several measures of entanglement have been suggested and
investigated: the entanglement of
formation~\cite{Bennett1996,Cbennett1996}, the entanglement of
distillation~\cite{Bennett1996,Cbennett1996}, the relative
entropy~\cite{Vedral1996,Vedral2002}, the robustness of
entanglement~\cite{Vidal1999}, and others.  In this paper we explore
a measure of entanglement called the {\it I-concurrence}, which was
introduced by Hill and Wootters \cite{Hill1997,Wootters1998} for
pairs of qubits and generalized to arbitrary bipartite systems by
Rungta {\em et al.}\ \cite{Rungta2001}.  Before turning to the
I-concurrence, we summarize here relevant results from entanglement
theory which will facilitate the discussion of I-concurrence.

Work on entanglement has served to identify certain {\em a priori\/}
axioms for a good measure of entanglement~\cite{Vidal2000}.
Entanglement characterizes intrinsically quantum-mechanical
correlations between quantum systems.  If $E(\rho)$ denotes an amount
of entanglement in the joint state $\rho$ of several quantum systems,
a fundamental requirement for $E$ to be a good measure of
entanglement is the following:
\begin{itemize}
\item[1.]  $E(\rho)$ does not increase, on average, under local
operations and classical communication (LOCC).  Such a measure is
called an {\em entanglement monotone\/} \cite{Vidal2000}.
\end{itemize}
An entanglement monotone must remain the same under the action of
reversible LOCC.  Hence, it is invariant under local unitary
transformations since these can be locally reversed.  Moreover, any
state $\rho$ can be converted into any separable state using LOCC;
therefore, an entanglement monotone takes a common minimum value for
all separable states, which can always be adjusted to
zero~\cite{Vidal2000}. Thus we can impose an additional,
nonnegativity requirement for a good entanglement measure:
\begin{itemize}
\item[2.] $E(\rho)\geq 0$ and goes to zero if and only if $\rho$ is a
separable state.
\end{itemize}

Henceforth, we restrict our discussion to bipartite quantum systems.
In three or more Hilbert-space dimensions, more than one measure is
required to characterize the entanglement of bipartite states
\cite{Vidal1999,Vidal2000,Vidal1999b}: to specify completely a joint
pure state of a $d\times d$ system, up to local actions, requires
$d-1$ independent Schmidt coefficients. Vidal~\cite{Vidal2000} showed
how to construct an infinite family of entanglement monotones for
bipartite systems.  This family of entanglement measures, denoted by
$\nu(\rho)$, is characterized by the following conditions.
\begin{itemize}
\item[(i)]For a pure state $|\Psi\rangle\langle\Psi|$, the measure
is a function of the marginal density operator $\rho_A={\rm
tr}_B(|\Psi\rangle\langle\Psi|)$, i.e., $\nu(\Psi)=f(\rho_A)$, where
the function $f$ has the following properties: (a)~invariance
under unitary transformations, i.e.,
$f(U\rho_AU^\dagger)=f(\rho_A)$, which implies that $f(\rho_A)$ is a
function of the eigenvalues of $\rho_A$, and (b)~concavity, i.e.,
$f(\lambda_1\rho_1+\lambda_2\rho_2)\geq
\lambda_1f(\rho_1)+\lambda_2f(\rho_2)$, where $\lambda_1$,
$\lambda_2\geq 0$ and $\lambda_1+\lambda_2=1$.

\item[(ii)] For a mixed state $\rho$, the measure is defined to
be the {\em convex-roof\/} extension of the pure-state measure, i.e.,
the minimum average value of the measure over all ensemble
decompositions of $\rho$:
\begin{equation}
\nu(\rho)\equiv\min_{\{p_j,|\Psi_j\rangle\}}
\Biggl\{\sum p_j\nu(\Psi_j)
\Biggm|\sum_jp_j|\Psi_j\rangle\langle\Psi_j|=\rho\Biggr\}\;.
\end{equation}
\end{itemize}
Vidal~\cite{Vidal2000} showed that if a measure belongs to the above
family, it satisfies property $1$ and thus is an entanglement
monotone.  The I-concurrence, as we show in Sec.~\ref{sec:properties},
belongs to this family.  Vidal also showed that when restricted to
pure states, any entanglement monotone satisfies condition~(i).

A privileged example of an entanglement monotone comes from choosing
the unitarily invariant concave function of condition~(i) to be the
von Neumann entropy, $f(\rho_A)=S(\rho_A)=-{\rm
tr}(\rho_A\log_2\rho_A)$.  This entanglement measure plays a special
role because of the asymptotic reversibility of the processes of
entanglement dilution and concentration for pure states, with von
Neumann entropy entering as the currency for these reversible
transformations \cite{Popescu1997}: $nS(\rho_A)$ Bell states can be
converted by LOCC to $n$ copies of $|\Psi\rangle$ in the limit as $n$
goes to infinity and vice versa \cite{Bennett1996}.  The
corresponding entanglement monotone for pure states,
$E_f(\Psi)=S(\rho_A)$, when extended to mixed states by the convex
roof, is called the {\em entanglement of formation\/}
\cite{Bennett1996,Cbennett1996}:
\begin{equation}
E_f(\rho)\equiv\min_{\{p_j,|\Psi_j\rangle\}}\Biggl\{\sum p_jE_f(\Psi_j)
\Biggm|\rho=\sum_jp_j|\Psi_j\rangle\langle\Psi_j|\Biggr\} \;.
\label{eq:eform}
\end{equation}
Unlike the situation for pure states, however, it is not known
whether the entanglement of formation gives the asymptotic cost in
Bell states for creating many copies of a mixed state $\rho$ using
LOCC.  This identification depends on whether $E_f(\rho)$ is additive
\cite{Wootters2001,Wootters1998}.  Though $E_f(\Psi)$ is additive on
pure states, this only implies that $E_f(\rho)$ is {\em
subadditive\/} on mixed states.  If $E_f(\rho)$ is strictly
subadditive, then some other entanglement monotone, given by the von
Neumann entropy for pure states, but less than $E_f(\rho)$ for mixed
states, quantifies the asymptotic cost of creating many copies of a
mixed state \cite{Wootters2001,Hayden2001}.

The special role of von Neumann entropy has been reconciled with the
existence of many other entanglement monotones by Vidal
\cite{Vidal2000} and Nielsen \cite{Nielsen2000}.  Vidal considered a
class of entanglement monotones that are additive on pure states and
showed that they have different asymptotic properties than the
entanglement of formation.  Nielsen formulated a precise property of
{\em asymptotic continuity\/} that is satisfied by the von Neumann
entropy and showed that this property is the key to relating
entanglement monotones to the asymptotic reversibility of
entanglement dilution and concentration that holds for pure states.
Indeed, Nielsen showed that any entanglement monotone that is
additive on pure states and that satisfies his asymptotic continuity
property is given, for pure states, by the von Neumann entropy of the
marginal density operator.  This result of Nielsen's leaves open
the question of the entanglement cost for mixed states.

Entanglement monotones other than the entanglement of formation,
though they are not related to asymptotic transformations, are
nonetheless important for characterizing LOCC transformations between
finite numbers of copies of states.  Indeed, they have been used to
quantify the probability and fidelity of exact and approximate LOCC
transformations between pure states
\cite{Vidal1999b,Nielsen1999,Jonathan1999a,Vidal2000b}.  In this
paper we investigate the properties of entanglement monotones related
to a particular entanglement measure called the {\em I-concurrence}.

Hill and Wootters \cite{Hill1997} introduced the {\it concurrence} as
a measure of entanglement for pairs of qubits.  The concurrence for a
pure state of two qubits is defined with the help of the qubit
spin-flip operator, and it is extended to mixed states as the convex
roof. Wootters \cite{Wootters2001,Wootters1998} went on to derive an
explicit formula for concurrence of an arbitrary joint state $\rho$
of two qubits in terms of the eigenvalues of
$\sqrt\rho\tilde\rho\sqrt\rho$, where $\tilde\rho$ is obtained by
spin flipping $\rho$, and showed how to calculate the corresponding
entanglement of formation from the concurrence.

Rungta {\it et al.}~\cite{Rungta2001} generalized the notion of
concurrence to pairs of quantum systems of arbitrary dimension.  This
generalized concurrence for a joint pure state $|\Psi\rangle$ of a
$d_A\times d_B$ system is simply related to the purity of the
marginal density operators:
\begin{equation}
C(\Psi)=\sqrt{2[1-{\rm tr}(\rho_A^2)]}=\sqrt{2[1-{\rm tr}(\rho_B^2)]}\;.
\label{eq:c}
\end{equation}
The generalized concurrence is known as the {\em I-concurrence\/}
because it is defined in terms of the universal-inverter
superoperator~\cite{Rungta2001}, which was shown to be a natural
generalization to higher dimensions of the spin flip for qubits. For
the purposes of this paper, we need not introduce the universal
inverter, simply taking Eq.~(\ref{eq:c}) as the definition of the
I-concurrence for pure states.  The I-concurrence is extended to
mixed states $\rho$ by the convex roof:
\begin{equation}
\label{eq:conc}
C(\rho)\equiv\min_{\{p_j,|\Psi_j\rangle\}}\Biggl\{\sum p_jC(\Psi_j)
\Biggm|\sum_jp_j|\Psi_j\rangle\langle\Psi_j|=\rho\Biggr\} \;.
\end{equation}

For pairs of qubits, Coffman, Kundu, and Wooters \cite{Coffman2000}
called $C^2(\rho)$ the tangle of the state $\rho$.  In this paper we
prefer to refer to $C^2(\rho)$ simply as the {\em squared
I-concurrence\/} and to reserve the term {\em tangle\/} for the
quantity obtained by extending $\tau(\Psi)\equiv C^2(\Psi)$ to mixed
states by the convex roof:
\begin{equation}
\label{eq:tangle}
\tau(\rho)\equiv\min_{\{p_j,|\Psi_j\rangle\}}\Biggl\{\sum p_jC^2(\Psi_j)
\Biggm|\sum_jp_j|\Psi_j\rangle\langle\Psi_j|=\rho\Biggr\} \;,
\end{equation}
Osborne \cite{Osborne2002} has calculated the tangle of rank-2
density operators in $2\times d$ systems.  As noted by Osborne, the
definition of tangle in Eq.~(\ref{eq:tangle}) does no violence to the
original usage of Coffman, Kundu, and Wootters, because
$\tau(\rho)=C^2(\rho)$ for bipartite qubit states, the reason being
that all pure states in the optimal ensemble decomposition have the
same concurrence.  Osborne found the tangle as defined in
Eq.~(\ref{eq:tangle}) to be a more natural quantity in his analysis
of rank-2 density operators, just as we find it to be more natural
for isotropic states in our analysis below.

The motivation for this paper is to show that, like the entanglement
of formation, the I-concurrence and tangle are entanglement monotones
(Sec.~\ref{sec:properties}) and then to provide explicit formulae for
the I-concurrence and tangle of isotropic states
(Sec.~\ref{sec:isotropic}).  For our analysis of isotropic states, we
use the formalism developed by Terhal and Vollbrecht
\cite{Terhal2000} to calculate the entanglement of formation for
isotropic states.  We find that the tangle for isotropic states is
closely related to the corresponding entanglement of formation
(Sec.~\ref{sec:conc}).

\section{Properties of I-concurrence and tangle}
\label{sec:properties}

Henceforth, we will omit the ``I'' when referring to the
I-concurrence.  Consider the Schmidt decomposition of an arbitrary
joint pure state $|\Psi\rangle$ of a $d_A\times d_B$ system:
\begin{equation}
\label{eq:psi}
|\Psi\rangle=\sum_j
\sqrt{\mu_j}|a_j\rangle\otimes|b_j\rangle=
\sum_j\sqrt{\mu_j}\,
U_A|e_j\rangle\otimes U_B|e_j\rangle\;,
\end{equation}
The squared Schmidt coefficients, $\mu_j$, are the eigenvalues of the
marginal density operators, $\rho_A$ and $\rho_B$, of the two systems,
and the vectors $|a_j\rangle$ and $|b_j\rangle$ make up the
orthonormal bases that diagonalize the marginal density operators.
These bases are connected to a fiducial orthonormal basis
$\{|e_j\rangle\}$ by unitary transformations $U_A$ and $U_B$. The
state $|\Psi\rangle$ is specified by its Schmidt vector
${\vec{\mu}}\equiv(\sqrt{\mu_1},\ldots,\sqrt{\mu_d})$ and the unitary
operators $U_A$ and $U_B$.

The tangle, or squared concurrence, of the pure state $|\Psi\rangle$
[see Eq.~(\ref{eq:c})] is given in terms of the Schmidt coefficients
by
\begin{equation}
\tau(\Psi)=C^2(\Psi)=
2\Biggl(1-\sum_j\mu_j^2\Biggr)=
4\sum_{j<k}\mu_j\mu_k\equiv
C^2({\vec{\mu}})\;.
\label{eq:Csquare}
\end{equation}
$C^2(\Psi)$ is conserved under by local unitary transformations
because it is a function only of the Schmidt coefficients
[property~(i)a].  It varies smoothly from $0$, for pure product
states, to $2(d-1)/d$, where $d\equiv\min(d_A,d_B)$, for maximally
entangled pure states.

That $C^2(\Psi)$ is a concave function of $\rho_A$ ($\rho_B$)
[property~(i)b] follows from the fact that $f(x)=-x^2$, $0\leq x\leq
1$, is a concave function, since for any concave function $f$, the
mapping $\rho\rightarrow{\rm tr}(f(\rho))$ is
concave~\cite{Wehrl1978}).  To see this, let
$\rho=\lambda_1\rho_1+\lambda_2\rho_2$, with $\lambda_1,\lambda_2\geq
0$ and $\lambda_1+\lambda_2=1$, and let $\rho=\sum_j
p_j|\phi_j\rangle\langle\phi_j|$, $\rho_1=\sum_k
q_k|\xi_k\rangle\langle\xi_k|$, and $\rho_2 =\sum_l
r_l|\eta_l\rangle\langle\eta_l|$ be eigendecompositions.  Then we
have
\begin{eqnarray}
{\rm tr}(f(\rho))
    &=& \sum_j f(\langle\phi_j|\rho|\phi_j\rangle)
    \nonumber \\
    &=& \sum_j f(\lambda_1\langle\phi_j|\rho_1|\phi_j\rangle
        +\lambda_2\langle\phi_j|\rho_2|\phi_j\rangle)
    \nonumber \\
    &\geq& \sum_j \lambda_1 f(\langle\phi_j|\rho_1|\phi_j\rangle)
        +\lambda_2 f(\langle\phi_j|\rho_2|\phi_j\rangle)
    \nonumber \\
    &=& \lambda_1\sum_j
        f\biggl(\sum_k|\langle\phi_j|\xi_k\rangle|^2 q_k\biggr)
        +\lambda_2\sum_j
        f\biggl(\sum_l|\langle\phi_j|\eta_l\rangle|^2 r_l\biggr)
    \nonumber \\
    &\geq& \lambda_1 \sum_{j,k}|\langle\phi_j|\xi_k\rangle|^2f(q_k)
        +\lambda_2 \sum_{j,l}|\langle\phi_k|\eta_l\rangle|^2f(r_l)
    \nonumber \\
    &=& \lambda_1\sum_k f(q_k)+\lambda_2\sum_l f(r_l)
    \nonumber \\
    &=& \lambda_1{\rm tr}(f(\rho_1))+\lambda_2{\rm tr}(f(\rho_2))\;,
\end{eqnarray}
where the two inequalities follow from the concavity of $f$. Since
adding a constant and multiplying by a positive constant doesn't
change concavity, $C^2(\Psi)$ is a concave function of $\rho_A$.
Since $C^2(\Psi)$ satisfies property~(i), its
extension~(\ref{eq:tangle}) by the convex roof to give the tangle
$\tau(\rho)$ of mixed states is an entanglement monotone.

The square root being an increasing concave function, it preserves
concavity.  Thus the concurrence $C(\Psi)$ is also a concave function
of $\rho_A$, and its extension~(\ref{eq:conc}) to give the
concurrence $C(\rho)$ of mixed states is an entanglement monotone.
Notice that since $C(\Psi)$ is zero if and only if $|\Psi\rangle$ is
a pure product state, $\tau(\rho)$ and $C(\rho)$ are zero if and only
if $\rho$ is separable.

From the properties of an entanglement monotone or directly from the
convex-roof construction, it is clear that the concurrence $C(\rho)$
is a convex function of bipartite density operators.  Since the
square is an increasing convex function, it preserves convexity, thus
making the squared concurrence, $C^2(\rho)$, a convex function of
bipartite density operators.  As a convex function that agrees with
the convex roof $\tau(\rho)$ on pure states, the squared concurrence
is guaranteed to satisfy $C^2(\rho)\le\tau(\rho)$.  We are unable to
say whether $C^2(\rho)$ is itself an entanglement monotone, although
it seems unlikely that it satisfies the property of not increasing
under local measurements.

\section{Tangle and concurrence of isotropic states}
\label{sec:isotropic}

\subsection{Isotropic states}

In this section we derive the tangle and concurrence of isotropic
states.  At the end of the section, we compare our results with the
those for the entanglement of formation of isotropic states obtained
by Terhal and Vollbrecht~\cite{Terhal2000}.  The tangle of isotropic
states shares important features with the entanglement of formation,
but the concurrence is significantly different.

Isotropic states are a class of mixed states for $d\times d$ systems
(two qudits); they are convex mixtures of the maximally mixed state,
$I_{d^2}={I}{\otimes}{I}/{d^2}$, with a maximally entangled state,
\begin{equation}
\label{eq:maxant}
|\Psi^+\rangle\equiv
{1\over\sqrt d}\sum_{j=1}^{d}|e_j\rangle\otimes|e_j\rangle\;.
\end{equation}
Such mixtures can be expressed as
\begin{equation}
\label{eq:epmaxent}
\rho_F=
{{1-F}\over d^2-1}\left(I-|\Psi^+\rangle\langle\Psi^+|\right)
+F|\Psi^+\rangle\langle\Psi^+|\;,
\end{equation}
where $F=\langle\Psi^+|\rho_F|\Psi^+\rangle$, satisfying $0\le
F\le1$, is the {\em fidelity\/} of $\rho_F$ and $|\Psi^+\rangle$.
These states were shown to be separable for $F\leq
1/d$~\cite{Rungta2001b,Horodecki1999}.

The isotropic states are special in the sense that they are invariant
under the action of the twirling superoperator ${\cal T}$
\cite{Horodecki1999}:
\begin{equation}
{\cal T}(\rho_F)=\int dU\,U\otimes U^*\rho_F U^\dagger\otimes
{U^*}^\dagger=\rho_F\;.
\end{equation}
Indeed, the twirling superoperator reduces any two-qudit state $\rho$
to an isotropic state
\begin{equation}
{\cal T}(\rho)=\rho_{F(\rho)}\;,
\end{equation}
where ${F(\rho)}=\langle\Psi^+|\rho|\Psi^+\rangle$ is the fidelity of
$\rho$ and $|\Psi^+\rangle$.  Twirling the pure state $|\Psi\rangle$
of Eq.~(\ref{eq:psi}) yields an isotropic state
\begin{equation}
\label{eq:tpsi}
{\cal T}(|\Psi\rangle\langle\Psi|)
=\rho_{F({\vec{\mu}},V)}^{\phantom{\dagger}}\;,
\end{equation}
where the fidelity is given by
\begin{eqnarray}
F({\vec{\mu}},V)
    &=& |\langle\Psi^+|\Psi\rangle|^2 \nonumber \\
    &=& {1\over d}\left|\sum_{j,k}\sqrt{\mu_k}
        \langle e_j|U_A|e_k\rangle\langle e_j|U_B|e_k\rangle\right|^2
    \nonumber \\
    &=& {1\over d}\left|\sum_{j,k}
        \sqrt{\mu_k}(U_A)_{jk}(U_B)_{jk}\right|^2 \nonumber \\
    &=& {1\over d}\left|\sum_{k}\sqrt{\mu_k}
        (U_A^TU_B)_{kk}\right|^2 \nonumber \\
    &=& {1\over d}\left|\sum_{k}\sqrt{\mu_k}\,V_{kk}\right|^2\;,
\end{eqnarray}
$V=U_BU_A^T$ being a unitary matrix.  It is easy to see that
\begin{equation}
\label{eq:fprime}
F({\vec{\mu}},V)\leq F({\vec{\mu}},I)\;,
\end{equation}
since $|V_{kk}|\leq 1$; equality holds if and only if
$V=I\exp(i\delta)$.

\subsection{Reduction to a single extremization}

We turn now to finding the tangle and concurrence of isotropic
states, using the technique developed for entanglement of formation
by Terhal and Vollbrecht \cite{Terhal2000}.  We proceed with the
analysis in terms of the tangle.  The reader should note that nothing
would change in the analysis if we replaced pure-state tangle
$C^2(\Psi)$ with pure-state concurrence $C(\Psi)$, thus analyzing
$C(\rho)$ instead of $\tau(\rho)$, till we get to
Sec.~\ref{sec:results}. At that point, we explicitly note the
differences between the results for tangle and concurrence.

The tangle of an arbitrary bipartite state $\rho$ satisfies the
following inequality,
\begin{eqnarray}
\tau(\rho) &\equiv&
\min_{\{p_j,|\Psi_j\rangle\}}
\Biggl\{\sum_j p_j
C^2(\Psi_j)\Biggm|\sum_j p_j|\Psi_j\rangle\langle\Psi_j|=\rho\Biggr\}
\label{eq:firsteq}\\
&\geq&  \min_{\{p_j,{\vec{\mu}}_j,V_j\}}
\Biggl\{\sum_j p_j C^2({\vec{\mu}}_j)\Biggm|
\sum_j p_jF({\vec{\mu}}_j,V_j)=F(\rho)\Biggr\}\;.
\label{eq:firstineq}
\end{eqnarray}
The inequality in Eq.~(\ref{eq:firstineq}) follows because an optimal
decomposition of $\rho$, i.e., one which achieves the minimum in
Eq.~(\ref{eq:firsteq}), automatically generates a set
$\{p_j,{\vec{\mu}}_j,V_j\}$ that satisfies the constraint in
Eq.~(\ref{eq:firstineq}).  In contrast, a set
$\{p_j,{\vec{\mu}}_j,V_j\}$ that achieves the minimum in
Eq.~(\ref{eq:firstineq}) generally does not generate an ensemble
decomposition of $\rho$.  Additional simplification is achieved by
splitting the minimization in Eq.~(\ref{eq:firstineq}) into two
parts, i.e.,
\begin{equation}
\label{eq:secondeq}
\tau(\rho)\geq\min_{\{p_j,V_j,F_j\}}
\Biggl\{\sum_j p_j
C^2_{V_j}(F_j)\Biggm|\sum_j p_jF_j=F(\rho)\Biggr\}
\equiv\tau\bigl(F(\rho)\bigr)\;,
\end{equation}
with
\begin{equation}
\label{eq:rv}
C^2_V(F)\equiv
\min_{\vec{\mu}}\Biggl\{C^2({\vec{\mu}})\Biggm|
F({\vec{\mu}},V)
\equiv{1\over d}\biggl|\sum_k V_{kk}\sqrt{\mu_k}\,\biggr|^2=F
\Biggr\}\;.
\end{equation}
The function $\tau(F)$ defined in Eq.~(\ref{eq:secondeq}) is a
function of the single parameter $F=F(\rho)$.

We can reduce the minimization problem further by noting that if
$\vec\nu$ is the vector of Schmidt coefficients that provides the
minimum for $C_V^2(F)$, then $F'\equiv F(\vec\nu,I)\ge
F(\vec\nu,V)=F$ [Eq.~(\ref{eq:fprime})] and
\begin{equation}
\label{eq:rvf}
C^2(F')\equiv C_I^2(F')\le C^2(\vec\nu)=C_V^2(F)\;.
\end{equation}
We find an explicit expression for $C^2(F)$ below and show that it is
monotonically increasing, from which it follows that
\begin{equation}
C_V^2(F)\ge C^2(F')\ge C^2(F)\;.
\label{eq:CCC}
\end{equation}
Applying Eq.~(\ref{eq:CCC}) to Eqs.~(\ref{eq:secondeq}) and
(\ref{eq:rv}) yields
\begin{eqnarray}
&&\tau(F)=
\min_{\{p_j,F_j\}}\Biggl\{\sum_j p_j
C^2(F_j)\Biggm|\sum_j p_jF_j=F\Biggr\}\;,
\label{eq:ineq1}\\
&&C^2(F)=\min_{\vec{\mu}}\Biggl\{C^2({\vec{\mu}})\Biggm|
{1\over d}\biggl(\sum_k\sqrt{\mu_k}\,\biggr)^2=F
\Biggr\}\;.
\label{eq:CCF}
\end{eqnarray}
Notice that the inequality $C^2(F')\ge C^2(F)$ in
Eq.~(\ref{eq:ineq1}) requires us to assume that $F\ge1/d$, since the
minimum that defines $C^2(F)$ does not exist for $F<1/d$.  This is
not a problem for the analysis of isotropic states, since isotropic
states with $F\le1/d$ are separable, having $\tau(\rho_F)=0$ and
$C(\rho_F)=0$.  The function $\tau(F)$ is by definition a convex
function of $F$.  Indeed, $\tau(F)$ can be defined as the largest
convex function that is bounded above by $C^2(F)$; it is given either
by $C^2(F)$ or by straight-line segments that connect points on the
graph of $C^2(F)$ and lie beneath $C^2(F)$.

We have shown that the tangle for any bipartite pure state is bounded
below by $\tau(\rho)\ge\tau\bigl(F(\rho)\bigr)$.  Now we use the
twirling superoperator to show that isotropic states achieve this
lower bound. To do so, consider an isotropic state $\rho_F$.  Let
$\{p_j,F_j,\vec{\mu}_j\}$ be a set that achieves the minimum in
Eqs.~(\ref{eq:ineq1}) and (\ref{eq:CCF}) with this value of $F$, and
let $\{|\Psi_j\rangle\}$ be states constructed from the Schmidt
vectors $\{\vec{\mu}_j\}$ with $V_j=I$.  The state formed from
this ensemble,
\begin{equation}
\rho=\sum_jp_j|\Psi_j\rangle\langle\Psi_j|\;,
\label{eq:optimum}
\end{equation}
has a tangle that satisfies
$\tau(\rho)\le\sum_jp_jC^2(\Psi_j)=\tau(F)$. Twirling $\rho$ gives
\begin{equation}
\label{eq:tbothsides}
{\cal T}(\rho)=
\sum_jp_j{\cal T}(|\Psi_j\rangle\langle\Psi_j|)
=\sum_jp_j\rho_{F_j}^{\phantom{\dagger}}
=\rho_F^{\phantom{\dagger}}\;,
\end{equation}
where we have made use of Eq.~(\ref{eq:tpsi}) to write ${\cal
T}(|\Psi_j\rangle\langle\Psi_j|)=\rho_{F_j}^{\phantom{\dagger}}$.
Since the local operations involved in twirling cannot increase the
tangle, we have $\tau(\rho)\ge\tau(\rho_F)$, from which follows the
upper bound $\tau(\rho_F)\le\tau(F)$.

Combined with the lower bound $\tau(\rho_F)\ge\tau(F)$, this shows
that the tangle of an isotropic state $\rho_F$ is given by the
function $\tau(F)$.  We have thus reduced the problem of finding the
tangle of an isotropic state to a single minimization, that of
finding the function $C^2(F)$, from which $\tau(\rho_F)=\tau(F)$ can
be constructed as described above.  If we follow through the steps of
this subsection for the concurrence, instead of the tangle, we find
that the concurrence of an isotropic state, $C(\rho_F)$, is the
largest convex function that is bounded above by $C(F)$.

\subsection{The extremization}
\label{sec:finalanswer}

Following the method of Terhal and Vollbrecht~\cite{Terhal2000}, we
calculate $C^2(F)$ using the method of Lagrange multipliers; i.e., we
minimize $C^2({\vec{\mu}})$ [Eq.~(\ref{eq:Csquare})] subject to the
constraints
\begin{eqnarray}
\sum_j\mu_j&=&1\;,\\
\sum_j \sqrt{\mu_j}&=&\sqrt{Fd}\;,
\end{eqnarray}
with $Fd\ge1$.  In doing the extremization, we allow for the
possibility that the minimum might not have all nonzero Schmidt
coefficients by explicitly considering all the cases where the number
of nonzero Schmidt coefficients varies from 1 to $d$.  The condition
for an extremum is
\begin{equation}
(\sqrt\mu_j)^3+\lambda_1\sqrt\mu_j+\lambda_2=0\;,
\end{equation}
where $\lambda_1$ and $\lambda_2$ are Lagrange multipliers.  The
three solutions of this cubic equation for $\sqrt{\mu_j}$ sum to
zero, so there are at most two real, positive solutions.  Letting
$\gamma$ and $\delta$ denote these two positive solutions, with
$\gamma>\delta$, the possible Schmidt vectors $\vec\mu$ have
coefficients
\begin{eqnarray}
\mu_j&=&\cases{
\gamma^2\;,&$j=1,\ldots,n$,\cr
\delta^2\;,&$j=n+1,\ldots,n+m$,\cr
0\;,&$j=n+m+1,\ldots,d$,}
\label{eq:coeff}
\end{eqnarray}
where $n+m\leq d$ and $n\ge1$.  The corresponding extrema of
$C^2(\vec\mu)$ are
\begin{equation}
\label{eq:rnm}
C^2_{nm}(F)=2\left(1-n\gamma^4-m\delta^4\right)\;,
\end{equation}
with the constraints
\begin{eqnarray}
\label{eq:cons}
n\gamma^2+m\delta^2&=&1\;,\nonumber\\
n\gamma+m\delta&=&\sqrt{Fd}\;.
\end{eqnarray}

Solving Eqs.~(\ref{eq:cons}), we obtain
\begin{eqnarray}
\label{eq:gam}
\gamma_{nm}^{\pm}(F)&=&
{n{\sqrt{Fd}\pm\sqrt{nm(n+m-Fd)}}\over n(n+m)}\;,\\
\label{eq:del}
\delta_{nm}^{\pm}(F)&=&
{\sqrt{Fd}-n\gamma_{nm}^\pm\over m}=
{m{\sqrt{Fd}\mp\sqrt{nm(n+m-Fd)}}\over m(n+m)}\;.
\end{eqnarray}
The relation $\delta_{nm}^\pm(F)=\gamma_{mn}^\mp(F)$ means that as we
vary $n$ and $m$ over all possible values, we need only consider the
upper sign; henceforth we drop the signs, using always the upper
sign. For the expressions~(\ref{eq:gam}) and (\ref{eq:del}) to give
real values, the quantity inside the square root must be nonnegative,
which implies that $Fd\le n+m$; moreover, in order that $\delta_{nm}$
be nonnegative, we must have $Fd\ge n$.  It is easy to see that
$\delta_{nm}(F)\le\sqrt{Fd}/(n+m)\le\gamma_{nm}(F)$, confirming that
the choice of the upper sign corresponds to our assumption that
$\gamma>\delta$. Notice also that $n=0$ is ill defined, as expected,
thus requiring $n\ge1$.

The function $C^2(F)$ we seek is the minimum of $C^2_{nm}(F)$ over
all choices of $n$ and $m$.  We can perform the minimization
explicitly by regarding $n$ and $m$ as continuous variables (here our
treatment departs from that of Terhal and Vollbrecht).  The task is
to minimize $C_{nm}^2(F)$ on the parallelogram defined by $1\le n\le
Fd$ and $Fd\le n+m\le d$.  Notice that the parallelogram collapses to
a line when $Fd=1$, i.e., at the separability boundary. As already
noted, within the parallelogram we have
$\gamma_{nm}(F)\ge\delta_{nm}(F)\ge0$;
$\gamma_{nm}(F)=\delta_{nm}(F)$ if and only if $n+m=Fd$, and
$\delta_{nm}(F)=0$ if and only if $n=Fd$.  We first calculate the
derivatives of $\gamma_{nm}(F)$ and $\delta_{nm}(F)$ with respect to
$n$ and $m$ by differentiating the constraints~(\ref{eq:cons}):
\begin{eqnarray}
{\partial\gamma\over\partial n}
&=&{1\over2n}{2\gamma\delta-\gamma^2\over\gamma-\delta}\;,\nonumber\\
{\partial\delta\over\partial n}
&=&-{1\over2m}{\gamma^2\over\gamma-\delta}\;,\nonumber\\
{\partial\delta\over\partial m}
&=&-{1\over2m}{2\gamma\delta-\delta^2\over\gamma-\delta}\;,\nonumber\\
{\partial\gamma\over\partial m}
&=&{1\over2n}{\delta^2\over\gamma-\delta}\;.
\label{eq:partials}
\end{eqnarray}
These can be used in Eq.~(\ref{eq:rnm}) to calculate the partial
derivatives of $C_{nm}^2(F)$:
\begin{eqnarray}
\label{eq:partialCn}
{\partial C^2\over\partial n}&=&
2\gamma^2[\gamma^2-2\delta(\gamma+\delta)]\;,\\
{\partial C^2\over\partial m}&=&
2\delta^2[\delta^2-2\gamma(\gamma+\delta)]\le-6\delta^4\le0\;.
\label{eq:partialCm}
\end{eqnarray}
It is useful to introduce co\"ordinates $u\equiv m-n$ and $v\equiv
m+n$, which correspond to motion parallel to and perpendicular to the
$m+n={\rm constant}$ boundaries of the parallelogram.  The derivative
of $C^2$ with respect to $u$ is
\begin{equation}
\label{eq:partialCu}
{\partial C^2\over\partial u}=
-{1\over2}(\gamma+\delta)(\gamma-\delta)^3\le0\;.
\end{equation}

The inequalities in Eqs.~(\ref{eq:partialCm}) and
(\ref{eq:partialCu}) follow immediately from the fact that
$\gamma\ge\delta\ge0$ within the parallelogram.  It should be clear
that $\partial C_{nm}^2/\partial m$ is strictly negative within the
parallelogram and $\partial C_{nm}^2/\partial u$ is strictly negative
except on the boundary $m+n=Fd$, where it is zero.  These results
mean that the minimum of $C_{nm}^2(F)$ occurs at the vertex $n=1$,
$m=d-1$, thus giving
\begin{equation}
C^2(F)=C_{1,d-1}^2(F)
=2\!\left(1-\gamma_{1,d-1}^4-(d-1)\delta_{1,d-1}^4\right)\;.
\end{equation}
Here
\begin{eqnarray}
\gamma_{1,d-1}(F)&=&
\sqrt{F/d}\left(1+w\sqrt{d-1}\right)\;,\\
\delta_{1,d-1}(F)&=&
\sqrt{F/d}\left(1-w/\sqrt{d-1}\right)\;,
\end{eqnarray}
with $w\equiv\sqrt{(1-F)/F}$.  Henceforth we omit the subscripts that
specify the case $n=1$ and $m=d-1$, this being the only case of
interest.

We need to confirm that $C^2(F)$ is monotonically increasing.
Differentiating the constraints~(\ref{eq:cons}) gives
\begin{eqnarray}
{\partial\gamma\over\partial F}&=&
-{1\over2}\sqrt{{d\over F}}{\delta\over\gamma-\delta}\;,\\
(d-1){\partial\delta\over\partial F}&=&
{1\over2}\sqrt{{d\over F}}{\gamma\over\gamma-\delta}\;,
\end{eqnarray}
from which we calculate
\begin{equation}
{\partial C^2\over\partial F}=
4\sqrt{{d\over F}}\gamma\delta(\gamma+\delta)\ge0\;,
\end{equation}
equality holding if and only if $\delta=0$, which requires $Fd=1$,
i.e., the separability boundary.  We conclude that $C^2(F)$ [and
$C(F)$] are monotonically increasing for $1/d\le F\le1$, as promised.
It will be useful below to write $\partial C^2/\partial F$ explicitly
in terms of $F$ and $d$:
\begin{equation}
{\partial C^2\over\partial F}=
8{F\over d}
\left(1+w\sqrt{d-1}\right)
\left(1-{w\over\sqrt{d-1}}\right)
\left[1+w{1\over2}\left(\sqrt{d-1}-{1\over\sqrt{d-1}}\right)\right]\;.
\label{eq:derivativeCsquared}
\end{equation}

\subsection{Results}
\label{sec:results}

\subsubsection{Two qubits}

The case of two qubits ($d=2$) is special, so we discuss it
separately.  For two qubits, there is one extremum,
$C^2(F)=C^2_{11}(F)=(2F-1)^2$, $1/2\leq F\leq 1$.  Since $C^2(F)$ is a
convex function of $F$, it follows that the tangle of an isotropic
state $\rho_F$ is given by
\begin{equation}
\tau(\rho_F)=\cases{%
    0\;,&$0\le F\le1/2$,\cr
    (1-2F)^2\;,&$1/2\le F\le1$,}
\end{equation}
and that the pure states in an optimal ensemble decomposition for the
tangle all have the same tangle.  The same conclusions hold for the
concurrence, $C(\rho_F)=C(F)=1-2F$, $1/2\le F\le 1$, which agrees
with the general expression derived by Wootters~\cite{Wootters1998}.

\begin{figure}[h]
\begin{center}
\includegraphics[width=8.5cm]{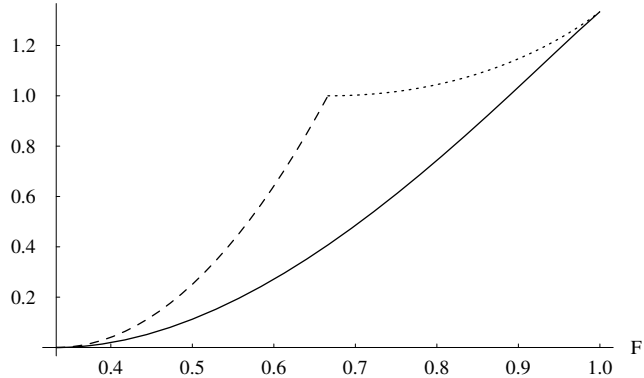}
\end{center}
\vspace{12pt}
\caption{Plots of $C^2_{11}(F)$ (dashed line),
$C^2_{21}(F)$ (dotted line), and $C^2_{12}(F)=C^2(F)$ (solid line)
for $d=3$.}
\label{fig1}
\end{figure}

\subsubsection{Two qutrits}

We consider the case of two qutrits ($d=3$) separately as an
illustration of what happens in the general qudit case.  Of the three
extrema, $C^2_{11}(F)$, $C^2_{21}(F)$, and $C^2_{12}(F)$, we already
know that the minimum is given by $C^2_{12}(F)$, i.e.,
$C^2(F)=C^2_{12}(F)$, $1/3\leq F\leq 1$.  This fact can also be
seen from Fig.~\ref{fig1}, where we have plotted the three extrema.

\begin{figure}[h]
\begin{center}
\includegraphics[width=8.5cm]{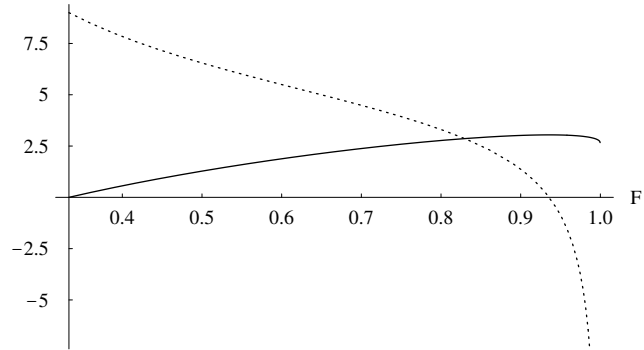}
\end{center}
\vspace{12pt}
\caption{First (solid line) and second (dotted line) derivatives of
$C^2(F)=C^2_{12}(F)$ for $d=3$.}
\label{fig2}
\end{figure}

\begin{figure}[ht]
\begin{center}
\includegraphics[width=8.5cm]{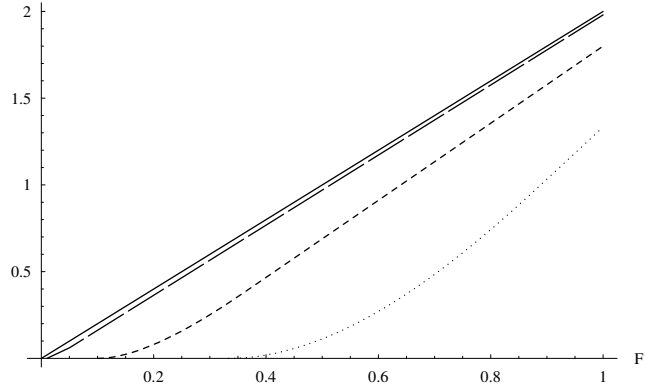}
\end{center}
\vspace{12pt}
\caption{Tangle $\tau(\rho_F)$ for $d=3$ (dotted), $d=10$
(short-dashed), $d=100$ (long-dashed), and $d=10\,000$ (solid). The
solid line is indistinguishable from the asymptotic tangle.}
\label{fig3}
\end{figure}

To calculate $\tau(\rho_F)$, it is necessary to analyze the behavior
of $C^2(F)$.  In Fig.~\ref{fig2}, we plot the first and second
derivatives of $C^2(F)$.  The first derivative confirms that $C^2(F)$
is monotonically increasing.  The second derivative changes sign from
positive to negative; $C^2(F)$ changes from convex to concave where
the second derivative vanishes.  The tangle $\tau(\rho_F)$ is the
largest convex function bounded above by $C^2(F)$, which is
constructed in the following way.  Find the (unique) line tangent to
$C^2(F)$ that passes through the point $(F=1,C^2=4/3)$; for $F$
smaller than the tangent point, the tangle is given by $C^2(F)$, but
for $F$ larger than the tangent point, the tangle is given by the
line.  The tangent point is found by solving the equation $\partial
C^2/\partial F=[4/3-C^2(F)]/(1-F)$, which gives $F=8/9$.  The slope
of the line is $(\partial C^2/\partial F)|_{F=8/9}=3$, and the tangle
at the tangent point is $C^2(8/9)=1$. Thus the tangle for $d=3$ is
\begin{equation}
\tau(\rho_F)=
\cases{0\;,&$F\leq1/3$,\cr
       C^2(F)\;,&$1/3\leq F\leq8/9$,\cr
       3(F-1)+4/3\;,&$8/9\leq F\leq 1$.}
\label{eq:d3}
\end{equation}
This function is plotted in Fig.~\ref{fig3}.

The behavior of $C^2(F)$ means that the pure states in an optimal
ensemble decomposition for the tangle all have the same tangle for
$F\le 8/9$, but have two values of tangle, $1$ and $4/3$ (maximal
entanglement), for $F>8/9$.

\begin{figure}[h]
\begin{center}
\includegraphics[width=8.5cm]{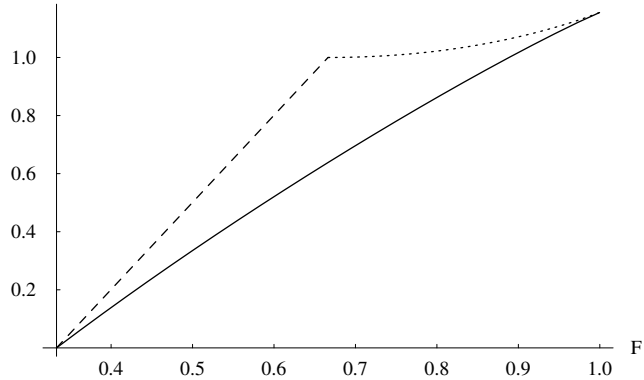}
\end{center}
\vspace{12pt}
\caption{Plots of $C_{11}(F)$ (dashed line), $C_{21}(F)$ (dotted
line), and $C_{12}(F)=C(F)$ (solid line) for $d=3$.}
\label{fig4}
\end{figure}

\begin{figure}[!]
\begin{center}
\includegraphics[width=8.5cm]{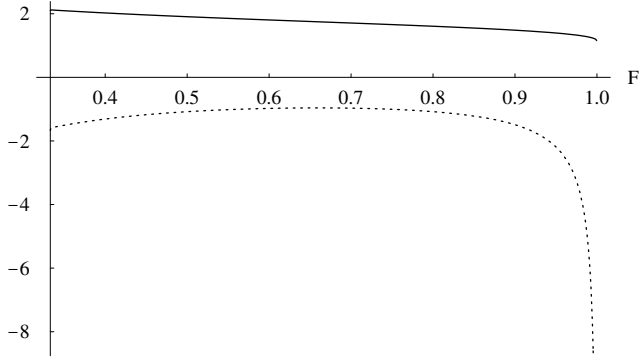}
\end{center}
\vspace{12pt}
\caption{%
First (solid line) and second (dotted line) derivatives of
$C(F)=C_{12}(F)$ for $d=3$.}
\label{fig5}
\end{figure}

Of the three extrema for the concurrence, $C_{11}(F)$, $C_{21}(F)$,
and $C_{12}(F)$, the minimum is given by $C_{12}(F)=C(F)$, a fact
confirmed by the plots in Fig.~\ref{fig4}.  In contrast to the
situation with the tangle, however, the second derivative of
$C(F)=C_{12}(F)$ is always negative (see Fig.~\ref{fig5}), which
means that $C(F)$ is concave.  Therefore the qutrit concurrence is
the straight line passing through the points $(F=1/3,C=0)$ and
$(F=1,C=\sqrt{4/3})$, i.e.,
\begin{equation}
C(\rho_F)=
\cases{0\;,&$F\leq1/3$,\cr
       \sqrt{3}(F-1/3)\;,&$1/3\leq F\leq 1$.
       }
\end{equation}
The concavity of $C(F)$ means that the pure states in an optimal
ensemble decomposition for the concurrence are either product states
or maximally entangled states.

\subsubsection{Two qudits}

For arbitrary $d$, we have already established that $C^2(F)=
C_{1,d-1}^2(F)$.  It turns out that for arbitrary $d\ge3$, $C^2(F)$
has the same behavior as for $d=3$; i.e., it changes from convex to
concave as $F$ increases. We get at this behavior by calculating the
second derivative of $C^2$ from Eq.~(\ref{eq:derivativeCsquared}):
\begin{equation}
{\partial^2 C^2\over\partial F^2}=
-{1\over2wF^2}{\partial\over\partial w}
\left({\partial C^2\over\partial F}\right)
=-{6\over w}{d-2\over d\sqrt{d-1}}
\left(1+{2\over 3}{d^2-8d+8\over(d-2)\sqrt{d-1}}w
-2w^2-{1\over 3}w^4\right)\;.
\label{eq:secondderivativeCsquared}
\end{equation}
Notice that for $d=2$ this expression simplifies to $\partial^2
C^2/\partial F^2=8$, as it should.  For $d\ge3$, the polynomial in
large parentheses in Eq.~(\ref{eq:secondderivativeCsquared}) clearly
has one root for positive $w$; it is easy to verify that this one
root occurs in the range of interest, i.e., between $F=1/d$
($w=\sqrt{d-1}$) and $F=1$ ($w=0$), showing that $C^2(F)$ changes
from convex to concave as $F$ increases.

We find the tangle $\tau(\rho_F)$ just as for $d=3$, i.e., by finding
the line tangent to $C^2(F)$ that passes through the point
$(F=1,C^2=2(d-1)/d)$.  Solving the equation $\partial C^2/\partial
F=[2(d-1)/d-C^2(F)]/(1-F)$ gives a tangent point at $F=4(d-1)/d^2$,
where the tangle and slope are $C^2=2(2d-3)/d(d-1)$ and
$\partial C^2/\partial F=2d/(d-1)$.  Thus the tangle is given by
\begin{equation}
\tau(\rho_F)=
\cases{0\;,&$F\leq1/d$,\cr
       C^2(F)\;,&$1/d\leq F\leq4(d-1)/d^2$,\cr
       2d(F-1)/(d-1)+2(d-1)/d\;,&$4(d-1)/d^2\leq F\leq 1$.}
\label{eq:tangleF}
\end{equation}
The tangle is plotted for a few representative values of $d$ in
Fig.~\ref{fig3}.  The pure states in an optimal ensemble
decomposition for the tangle all have the same tangle for $F\le
4(d-1)/d^2$, but have two values of tangle, $2(2d-3)/d(d-1)$ and
$2(d-1)/d$ (maximal entanglement), for $F>4(d-1)/d^2$.  As
$d\rightarrow\infty$, $\tau(\rho_F)\rightarrow 2F$ becomes a linear
function.

The concurrence also has the same behavior generally as for $d=3$. A
calculation of $\partial^2 C/\partial F^2$ using
Eqs.~(\ref{eq:derivativeCsquared}) and
(\ref{eq:secondderivativeCsquared}) shows that $C(F)$ is a concave
function on the range of interest.  Thus the concurrence of isotropic
states is a linear function between zero concurrence at the separability
boundary and maximal entanglement at $F=1$:
\begin{equation}
C(\rho_F)=
\cases{0\;,&$F\leq1/d$,\cr
       \sqrt{2d/(d-1)}(F-1/d)\;,&$1/d\leq F\leq 1$.}
\end{equation}
The pure states in an optimal ensemble decomposition for the
concurrence are either product states or maximally entangled states.
As $d\rightarrow\infty$, $C(\rho_F)\rightarrow {\sqrt 2}F$.

\section{Conclusion}
\label{sec:conc}

It is informative to conclude the paper by comparing our results for
the tangle of isotropic states with the results of Terhal and Vollbrecht
\cite{Terhal2000} for the entanglement of formation.

If one follows through the procedure of Terhal and
Vollbrecht~\cite{Terhal2000} for finding the entanglement of
formation, one finds that the function $C^2(F)$ is replaced by the
function
\begin{equation}
E(F)\equiv H_2(\gamma^2)+(1-\gamma^2)\log(d-1)\;,
\end{equation}
where $\gamma=\gamma_{1,d-1}$ and $H_2(x)=-x\log x-(1-x)\log(1-x)$ is
the binary entropy function.  The entanglement of formation,
$E(\rho_F)$, is given by the largest convex function that is bounded
above by $E(F)$.  Terhal and Vollbrecht conjecture that $E(F)$ has
the properties that we find here for $C^2(F)$, i.e., that $E(F)$ has
a single inflection point in the range $1/d\le F\le1$, changing from
convex to concave as $F$ increases.  Given this conjecture, the
entanglement of formation is found by finding the straight line
tangent to $E(F)$ that passes through the point $(F=1,E=\log d)$.
Remarkably the tangent point for $E(F)$ is the same as for $C^2(F)$,
which gives the entanglement of formation as
\begin{equation}
\label{eq:entropy}
E_f(\rho_F)=
\cases{%
    0\;,&$1/d\le F$,\cr
    E(F)\;,&$1/d\leq F\leq 4(d-1)/d^2$,\cr
    d\log(d-1)(F-1)/(d-2)+\log d\;,&$4(d-1)/d^2\leq F\leq 1$.}
\end{equation}
Asymptotically, as $d\rightarrow\infty$, the entanglement of
formation becomes $E_f(\rho_F)\rightarrow F\log d$.  The similarity of
the tangle and entanglement of formation must reflect a deep
connection between the two, at least for isotropic states, but we
have not been able to find a simple reason for this similarity.

\acknowledgments
This work was supported in part by Office of Naval Research Contract
No.~N00014-00-1-0578.


\begin{thebibliography}{99}

\bibitem{Einstein1935}
A.~Einstein, B.~Podolski, and N.~Rosen, Phys.\ Rev.\, {\bf 47},
777 (1935).

\bibitem{Bell1964}
J.~S.~Bell, Physics\, {\bf 1}, 195 (1964).

\bibitem{Wootters2001}
W.~K.~Wootters, Quant Inf.\ Comp. {\bf 1}, 27 (2001).

\bibitem{Lo1998}
{\em Introduction to Quantum Computation and Information}, edited
by H.-K.~Lo, S.~Popescu, and T.~Spiller (World Scientific,
Singapore, 1998).

\bibitem{Mnielsen2000}
M.~A. Nielsen and I.~L.~Chuang, {\em Quantum Computation and
Quantum Information} (Cambridge University Press, Cambridge,
England, 2000).

\bibitem{Bennett1996}
C.~H. Bennett, H.~J. Bernstein, S.~Popescu, and B.~Schumacher,
Phys.\ Rev.~A\, {\bf 53}, 2046 (1996).

\bibitem{Cbennett1996}
C.~H. Bennett, D.~P. DiVincenzo, J.~A. Smolin, and W.~K. Wootters,
Phys.\ Rev.\ A {\bf 54}, 3824 (1996).

\bibitem{Vedral1996}
V.~Vedral and M.~B.~Plenio, Phys.\ Rev.\ A {\bf 57}, 1619
(1996).

\bibitem{Vedral2002}
V.~Vedral, Rev.\ Mod.\ Phys.\ {\bf 74}, 197 (2002).

\bibitem{Vidal1999}
G.~Vidal and R.~Tarrach, Phys.\ Rev.\ A {\bf 59}, 141 (1999).

\bibitem{Hill1997}
S.~Hill and W.~K. Wootters, Phys.\ Rev.\ Lett.\ {\bf 78}, 5022
(1997).

\bibitem{Wootters1998}
W.~K. Wootters, Phys.\ Rev.\ Lett.\ {\bf 80}, 2245 (1998).

\bibitem{Rungta2001}
P.~Rungta, V.~Bu\v{z}ek, C.~M.~Caves, M.~Hillery, and G.~J. Milburn,
Phys.\ Rev.~A {\bf 64}, 042315 (2001).

\bibitem{Vidal2000}
G.~Vidal, J.\ Mod.\ Opt.\ {\bf 47}, 355 (2000).

\bibitem{Vidal1999b}
G.~Vidal, Phys.\ Rev.\ Lett. {\bf 83}, 1046 (1999).

\bibitem{Popescu1997}
S. Popescu and D. Rohrlich, Phys.\ Rev.~A {\bf 56}, R3319 (1997).

\bibitem{Hayden2001}
P. Hayden, M. Horodecki, and B. M. Terhal, J.\ Phys.\ A
{\bf 34}, 6891 (2001).

\bibitem{Nielsen2000}
M.~A. Nielsen, Phys.\ Rev.\ A {\bf 61}, 064301 (2000).

\bibitem{Nielsen1999}
M.~A. Nielsen, Phys.\ Rev.\ Lett.\ {\bf 83}, 436 (1999).

\bibitem{Jonathan1999a}
D.~Jonathan and M.~B.~Plenio, Phys.\ Rev.\ Lett. {\bf 83}, 1455 (1999).

\bibitem{Vidal2000b}
G.~Vidal and D.~Jonathan and M.~A. Nielsen, Phys.\ Rev.~A {\bf 62},
012304 (2000).

\bibitem{Coffman2000}
V.~Coffman, J.~Kundu, and W.~K. Wootters, Phys.\ Rev.~A {\bf 61},
052306 (2000).

\bibitem{Osborne2002}
T.~J.~Osborne, unpublished, {\tt arXiv.org e-print quant-ph/0203087}.

\bibitem{Terhal2000}
B.~M.~Terhal and K.~G.~H.~Vollbrect, Phys.\ Rev.\ Lett.\ {\bf 85},
2625 (2000).

\bibitem{Wehrl1978}
A.~Wehrl, Rev.\ Mod.\ Phys.\ {\bf 50}, 221 (1978).

\bibitem{Rungta2001b}
P.~Rungta, W.~J. Munro, K.~Nemoto, P.~Deuar, G.~J. Milburn, and C.~M.
Caves, in Directions in Quantum Optics: A Collection of Papers
Dedicated to the Memory of Dan Walls, edited by H.~J. Carmichael,
R.~J. Glauber, and M.~O. Scully (Springer, Berlin, 2001), p.~149.

\bibitem{Horodecki1999}
M.~Horodecki and P.~Horodecki, Phys.\ Rev.~A {\bf 59}, 4206
(1999).

\end{thebibliography}
\end{document}